\documentstyle[12pt]{article}
\begin{document}

\newcommand{\be}{\begin{equation}}
\newcommand{\ee}{\end{equation}}

\renewcommand{\thefootnote}{\fnsymbol{footnote}}
\parindent=3pc

\newcommand{\sheptitle}
{Supersymmetric Higgs Bosons at the Limit
\footnote{Talk presented at the HARC SUSY Workshop, Texas, April 1993.}}

\newcommand{\shepauthor}
{S. F. King,}

\newcommand{\shepaddress}
{Physics Department,\\University of Southampton,\\Southampton,\\SO9 5NH,\\U.K.}

\newcommand{\shepabstract}
{Using a combination of renormalisation group and effective
potential methods, we discuss the bound on the lightest CP-even
Higgs boson mass $m_h$ in the next-to-minimal supersymmetric standard model.
We find $m_h\leq 146,139,149 \ GeV$ for $m_t=90,140,190 \ GeV$.}

\begin{titlepage}
\hfill SHEP 92/93-12
\vspace{.4in}
\begin{center}
{\Huge{\bf \sheptitle}}
\bigskip \\ \shepauthor \\ {\it \shepaddress} \\ \vspace{.5in}
{\bf Abstract} \bigskip \end{center} \setcounter{page}{0}
\shepabstract
\end{titlepage}

\newpage

{\bf\noindent 1. Introduction}
\vglue 0.2cm
{\it\noindent 1.1. Triviality}
\vglue 0.1cm
\baselineskip=24pt

The question of the origin of electroweak mass is one
of the most urgent questions of present day particle physics.
The discovery of a particle which resembles the Higgs boson
of the minimal standard model, and the measurement of its mass,
will provide clues as to the nature of new physics beyond the standard model.
In this report we shall be concerned with the question
of how heavy the lightest neutral CP-even
supersymmetric Higgs boson,
$h^0$, can be within the framework of supersymmetric grand unified theories
(SUSY GUTs) \cite{1}.
In SUSY GUTs all the Yukawa couplings are constrained
to remain perturbative in the region
$M_{SUSY}\sim 1$ TeV to $M_{GUT}\sim 10^{16}$ GeV.
This constraint provides a maximum value at low energies for those
Yukawa couplings which are not asymptotically-free, and is
obtained from the renormalisation group (RG) equations
together with the boundary
conditions that the couplings become non-perturbative
at $M_{GUT}$ -- the so-called ``triviality limit''.
In the minimal supersymmetric standard model (MSSM)
\cite{2}, the triviality limits provide a useful bound on the top quark mass
$m_t$. The upper bound on the $h^0$ mass, $m_h$, in the MSSM,
including radiative corrections,
has recently been the subject of much discussion\cite{6,7,8,9,10,11,12}.
However the MSSM is not the most general
low energy manifestation of SUSY \hbox{GUTs.}
It is possible that SUSY GUTs give
rise to a low energy theory which contains an additional
gauge singlet field, the so called next-to-minimal supersymmetric
standard model (NMSSM) \cite{3,4,5}.
Here we shall concentrate on the question
of the upper-bound on $m_h$ in the NMSSM
which is obtained from triviality limits of Yukawa couplings.

\vglue 0.2cm
{\it \noindent 1.2. The NMSSM}
\vglue 0.1cm

The NMSSM provides an elegant extension of MSSM
which eliminates the $\mu$--problem, the appearance of a mass--scale in
the superpotential. The NMSSM differs from the MSSM by the presence of a
gauge singlet field whose vacuum expectation value (vev) plays the role
of the mass parameter $\mu$ in the MSSM, and is defined by the superpotential
\be
W = h_t Q H_2 t^c + \lambda N H_1 H_2 - \frac{1}{3} k N^3 + \dots,
\label{superpotential}
\ee
where the superfield $Q^T=(t_L,b_L)$ contains the left--handed top and
bottom quarks, and $t^c$ contains the charge conjugate of the
right--handed top quark; $H_1$ and $H_2$ are the usual Higgs doublet
superfields and $N$ is the Higgs gauge singlet superfield; the ellipsis
represents terms whose relatively small couplings play no role in our
analysis (in particular, we assume that the bottom quark Yukawa coupling,
$h_b$, satisfies $h_b \ll h_t$).
The fields $H_1^T=(H_1^0, H_1^-)$, $H_2^T=(H_2^+, H_2^0)$ and
$N$ develop vevs which may be assumed to be of the form \cite{10}
\be
<H_1> = \left( \begin{array}{c} \nu_1 \\ 0 \end{array} \right), \ \ \
<H_2> = \left( \begin{array}{c} 0 \\ \nu_2 \end{array} \right), \ \ \
<N> = x,
\ee
where $\nu_1$, $\nu_2$ and $x$ are real, and $\sqrt{\nu_1^2+\nu_2^2}
= \nu =174$ GeV. The low energy physical spectrum of the Higgs scalars
consists of 3 CP-even neutral states, 2 CP-odd neutral states, and
2 charged scalars. A third CP-odd state is a Goldstone mode which becomes
the longitudinal component of the $Z^0$, while a further two charged scalars
become those of the $W^\pm$'s.

An upper bound on the lightest neutral CP-even scalar $h^0$ in the NMSSM
may be obtained from the real symmetric
$3\times 3$ neutral scalar mass squared matrix,
by using the fact that the smallest eigenvalue of such a matrix
must be smaller than the smallest eigenvalue
of its upper $2\times 2$ block.
The resulting bound at tree-level is \cite{5}
\begin{equation}
{m_{h}}^2\leq {M_Z}^2 +
(\lambda ^2v^2-{M_Z}^2)\sin^22\beta.
\end{equation}
where $\tan \beta \equiv \frac{v_2}{v_1}$,
and $\lambda$ is regarded as a running parameter
evaluated at $M_{SUSY}$.
The upper bound on $m_h$ is determined by
the maximum value of $\lambda (M_{SUSY})$, henceforth
denoted $\lambda _{max}$.
The value of $\lambda _{max}$ is obtained
by solving the SUSY RG equations for the Yukawa couplings $h_t$, $\lambda$
and $k$ in the region
$M_{SUSY}=1$ TeV to $M_{GUT}=10^{16}$ GeV \cite{13,14}.
If the Yukawa couplings $h_t$, $\lambda$ and $k$
are all initially large at $M_{GUT}$ then they approach low energy
fixed point ratios \cite{13}. However,
if the boundary condition at $M_{GUT}$ is $\lambda \gg k$,
then larger values of $\lambda (M_{SUSY})$ can be achieved.
We have repeated the calculation
of ref.\cite{14} and found that
for $h_t(M_{SUSY})=0.5-1.0$,
${\lambda}_{max}=0.87-0.70$ and
for $h_t(M_{SUSY}) \rightarrow 1.06$,\footnote{ $h_t(M_{SUSY})\leq 1.06$ is
the triviality bound, which, together with
$m_t=h_t(m_t)v\sin \beta$, where $h_t(m_t)\leq 1.12$,
implies the bound $m_t \leq 195$ GeV.}
${\lambda}_{max}\rightarrow 0$ (with $k=0$ always).
Radiative corrections to the tree-level bound
in Eq.(2) have been
considered in refs.\cite{15,16,17}.
In ref.\cite{16} these
were estimated from a low energy RG analysis of the Higgs sector
of the model between $M_{SUSY}$ and a lower scale $\mu$,
assuming that only one Higgs boson has a mass below
$M_{SUSY}$.

In Section 2 \cite{elliott1}
we shall discuss a more general RG analysis
in which both Higgs doublets and the Higgs singlet may
be lighter than $M_{SUSY}$, making a simple approximation
of hard decoupling below $M_{SUSY}=1 TeV$ of the superpartners.
In Section 3 \cite{elliott2}
we shall go on to consider the more general case
in which the squarks have a more general mass spectrum, using
the effective potential approach.
Our numerical results for the bound are also presented
in this section.
Section 4 contains our concluding remarks.

\vglue 0.6cm
{\bf\noindent 2. Renormalisation Group Approach \cite{elliott1}}
\vglue 0.4cm
The basic assumption of this approach is that the squarks
are degenerate at $M_{SUSY}=1$ TeV while the Higgs bosons and top
quarks have masses $\mu \approx 150$ GeV.
The effective theory below $M_{SUSY}$ is just the standard model
with two light Higgs doublets and a light Higgs singlet.
Thus the Higgs potential at some low energy scale $\mu <M_{SUSY}$
is given by the general expression
\begin{eqnarray}
V_0(\mu) & = & \frac{1}{2} {\lambda}_1({H_1}^{\dagger}H_1)^2
+ \frac{1}{2} {\lambda}_2({H_2}^{\dagger}H_2)^2
+ ({\lambda}_3 +{\lambda}_4)({H_1}^{\dagger}H_1)({H_2}^{\dagger}H_2)
\nonumber \\
          & - & {\lambda}_4|{H_2}^{\dagger}H_1|^2
+ {\lambda}_5|N|^2|H_1|^2 + {\lambda}_6|N|^2|H_2|^2 \nonumber \\
          & + & {\lambda}_7({N^{\ast}}^2H_1H_2 + H.c.)
+ {\lambda}_8|N|^4 \nonumber \\
          & + & {m_1}^2|H_1|^2 + {m_2}^2|H_2|^2 + {m_3}^2|N|^2 \nonumber \\
          & - &  m_4(H_1H_2N + H.c.)- \frac{1}{3} m_5(N^3 + H.c.).
\end{eqnarray}
The running quartic couplings ${\lambda}_i(\mu)$
and the mass parameters $m_i(\mu)$ must satisfy the following boundary
conditions at $M_{SUSY}$
\begin{eqnarray}
{\lambda}_1 &=& {\lambda}_2=\frac{1}{4} ({g_2}^2 + {g_1}^2),\ \ \
{\lambda}_3=\frac{1}{4} ({g_2}^2 - {g_1}^2) \nonumber \\
{\lambda}_4 &=& {\lambda}^2-\frac{1}{2}{g_2}^2,\ \ \
{\lambda}_5={\lambda}_6={\lambda}^2, \ \ \
{\lambda}_7=-{\lambda}k, \ \ \ {\lambda}_8=k^2, \nonumber \\
        m_1 & = & m_{H_1}, \ \ \ m_2  =  m_{H_2}, \ \ \
m_3  =  m_N, \nonumber \\
        m_4 &=& {\lambda}A_{\lambda}, \ \ \ m_5=kA_k,
\end{eqnarray}
where $A_{\lambda}$ and $A_{k}$ are soft parameters
associated with the trilinear Higgs couplings in Eq.(1),
and we have deferred a discussion of squark effects to Section 3.
At energy scales $\mu$ below $M_{SUSY}$, the values of the
quartic couplings may be obtained by solving
the RG equations given in ref.\cite{elliott1}.
The minimisation conditions implied by
$\frac{ \partial V_{Higgs}}{\partial v_i}=0$ and
$\frac{ \partial V_{Higgs}}{\partial x}=0$
allow us to eliminate the low energy parameters
$m_1$, $m_2$, $m_3$. The remaining masses $m_4$ and $m_5$
are related to the parameters $A_{\lambda}$ and $A_k$ at $M_{SUSY}$
by Eq.(5). Below this scale we shall regard $m_4$ and $m_5$ as
free parameters.

The neutral CP-even (scalar) mass squared symmetric matrix in the basis
$1,2,3=H_1,H_2,N$  is
\begin{eqnarray}
M^2 & = &
\left( \begin{array}{ccc}
2\lambda_1 \nu_1^2 & 2(\lambda_3+\lambda_4) \nu_1 \nu_2 & 2\lambda_5 x \nu_1
\\
2(\lambda_3+\lambda_4) \nu_1 \nu_2 & 2\lambda_2 \nu_2^2 & 2\lambda_6 x \nu_2
\\
2\lambda_5 x \nu_1 & 2\lambda_6 x \nu_2 & 4\lambda_8 x^2 - m_5 x
\end{array} \right) \nonumber \\
 & + & \left( \begin{array}{ccc}
\tan \beta [m_4x-\lambda_7x^2] & - [m_4x-\lambda_7x^2] & -\frac{\nu_2}{x}
[m_4x-2\lambda_7x^2] \\
- [m_4x-\lambda_7x^2] & \cot \beta [m_4x-\lambda_7x^2] & -\frac{\nu_1}{x}
[m_4x-2\lambda_7x^2] \\
-\frac{\nu_2}{x} [m_4x-2\lambda_7x^2] & -\frac{\nu_1}{x} [m_4x-2\lambda_7x^2]
& \frac{\nu_1 \nu_2}{x^2} [m_4x]
\end{array} \right)
\end{eqnarray}
The neutral CP-odd (pseudoscalar) mass squared
symmetric matrix is
\be
\tilde{M}^2 =
\left( \begin{array}{ccc}
\tan \beta [m_4x-\lambda_7x^2] & [m_4x-\lambda_7x^2] & \frac{\nu_2}{x}
[m_4x+2\lambda_7x^2] \\
\mbox{} [m_4x-\lambda_7x^2] & \cot \beta [m_4x-\lambda_7x^2] & \frac{\nu_1}{x}
[m_4x+2\lambda_7x^2] \\
\frac{\nu_2}{x} [m_4x+2\lambda_7x^2] & \frac{\nu_1}{x} [m_4x+2\lambda_7x^2] &
3m_5x + \frac{\nu_1 \nu_2}{x^2} [m_4x-4\lambda_7x^2]
\end{array} \right).
\ee

In ref.\cite{elliott1} we obtained an upper bound
on the mass of the lightest neutral scalar Higgs boson $h^0$,
by using the fact that $m_h^2$ must not exceed
the lower eigenvalue of the upper $2 \times 2$ block matrix.
The resulting upper
bound is a complicated function of ${m_c}^2$.
It is easy to show that the bound reaches a maximum
asymptotically for $m_c\rightarrow \infty$ \cite{elliott1}.
It is easy to show that the bound is maximised
for $\lambda >{\lambda}_{max}$ and $k=0$,
and so we shall use $\lambda _{max}$ and $k=0$,
as in the case of the tree level bound discussed previously.
With this information in hand, a bound may be calculated
from the values of $\delta {\lambda}_i={\lambda}_i(\mu)-{\lambda}_i(M_{SUSY})$
obtained from numerically integrating the RG equations,
decoupling the top quark below its mass and choosing $\mu=150$ GeV.
For each value of $m_t$, we have determined the value of $h_t(M_{SUSY})$
and the corresponding value of ${\lambda}_{max}$ which maximises the bound
from a numerical analysis of
the triviality condition as discussed previously.
We defer a discussion of this bound until after squark effects have been
considered.

\vglue 0.4cm

{\bf \noindent 3. Squark Contributions \cite{elliott2}}

\vglue 0.4cm

In order the calculate the shifts in the mass--squared matrices due to
squarks, we employ the full one--loop effective potential. Much of our
analysis of the squark spectrum is similar to that of the MSSM in
Ref. \cite{6}, whose notation we follow closely.
A similar approach has also been followed by Ellwanger \cite{15}.
The one--loop effective potential is given by
\begin{eqnarray}
V_1(Q) & = & V_0(Q) + \Delta V_1(Q), \nonumber \\
\Delta V_1(Q) & = & \frac{1}{64\pi^2} \mbox{Str} {\cal M}^4 \left(
\ln \frac{{\cal M}^2}{Q^2} - \frac{3}{2} \right),
\end{eqnarray}
where $V_0(Q)$ is the tree--level potential at the arbitrary $\overline{MS}$
scale $Q$, $\mbox{Str}$ denotes the usual supertrace, and ${\cal M}^2$ is the
field--dependent mass--squared matrix. We shall take $Q = M_{SUSY} = 1$ TeV.
\footnote{This value of $Q$ is of course totally arbitrary. The value of
1 TeV is chosen to be consistent with our previous analysis. The results
are independent of $Q$ to one loop order.}
$V_0$ is restricted to the pure Higgs part of the tree--level potential.
Since the RG approach of Section 2 \cite{elliott1}
has already included the logarithmic effects of top
quark and Higgs boson loops we exclude these particles from the supertrace.
The logarithmic contributions to $\Delta V_1 (Q)$ from the top quark and
Higgs bosons may be absorbed into $V_0(Q)$ where $Q = 1$ TeV, to yield
$V_0(\mu)$ where $\mu = 150$ GeV, as in Eq.(4). It only remains to
calculate $\Delta V_1(Q)$, where $Q = 1$ TeV, involving squark contributions.

To calculate the shifts in the Higgs mass matrices as a result of squark
effects we require the field--dependent mass--squared matrices of the
squarks. Taking the contribution to the soft SUSY breaking potential
involving squarks to be
\be
\Delta V_{soft} = h_t A_t (\tilde{Q} H_2 \tilde{t^c} + h.c.) +
                  m_Q^2 | \tilde{Q} |^2 +m_T^2 | \tilde{t^c} |^2 +
                  m_B^2 | \tilde{b^c} |^2,
\ee
where $b^c$ contains the charge conjugate of the right--handed bottom
quark, and tildes denote the scalar components of the superfields,
together with the superpotential in Eq.(1), leads to the
field--dependent squark mass--squared matrix (ignoring contributions
proportional to gauge couplings and $h_b$)
\be
{\cal M}^2 = \left(
\begin{array}{cccc}
m_Q^2 + h_t^2 |H_2^0|^2 & \lambda h_t N H_1^0 + h_t A_t \bar{H_2^0} & -h_t^2
\bar{H_2^0} H_2^+ & 0 \\
\lambda h_t \bar{N} \bar{H_1^0} + h_t A_t H_2^0 & m_T^2 + h_t^2
(|H_2^0|^2+H_2^+H_2^-) & \lambda h_t \bar{N} H_1^+ - h_t A_t H_2^+ & 0 \\
-h_t^2 H_2^0 H_2^- & \lambda h_t N H_1^- - h_t A_t H_2^- & m_Q^2 + h_t^2 H_2^+
H_2^- & 0 \\
0 & 0 & 0 & m_B^2
\end{array} \right),
\label{squarkmatrix}
\ee
in the basis
$\{ \tilde{t_L},\bar{\tilde{t_R^c}},\tilde{b_L},\bar{\tilde{b_R^c}} \}$
where a bar denotes complex conjugation. Manifestly, one eigenvalue is
field--independent and may be discarded. Only the upper $2 \times 2$ submatrix
contributes to the CP-even and CP-odd mass--squared matrices, since the
charged fields have zero vevs in order not to break QED, whereas the
upper $3 \times 3$ submatrix contributes to the charged mass--squared
matrix. By differentiating $\Delta V_1$ once with respect to the fields we
may obtain the shifts in the minimisation conditions induced by squark loops,
and twice will yield the shifts in the mass matrices.
\footnote{In fact, this is only approximately true due to Higgs self--energy
corrections; these are expected to be small for the lightest Higgs bosons
\cite{11}.}

In the basis $\{ H_1, H_2, N \}$ we have the following mass-squared
matrices, after ensuring that the full one--loop potential is correctly
minimised. The couplings $\lambda_i$ are those obtained from the potential
in Eq.(4) renormalised at the scale $\mu=150$ GeV.
For notational simplicity we drop the tildes. Moreover, we work in the basis
of mass eigenstates $\{ t_1 , t_2 , b_1 , b_2 \}$ (though, since $h_b$, the
bottom quark Yukawa coupling, is taken to be zero, $b_2$ never contributes to
the mass--squared matrices), where the mass of the $t_1$ squark is
$m_{t_1}$, and so on for the other squarks. The CP-even (scalar)
mass--squared matrix is
\be
M^2_s = M^2 + \delta M^2,
\ee
where $M^2$ is given in Eq.(6), and

\be
\delta M^2 =
\left( \begin{array}{ccc}
\Delta_{11}^2 & \Delta_{12}^2 & \Delta_{13}^2 \\
\Delta_{12}^2 & \Delta_{22}^2 & \Delta_{23}^2 \\
\Delta_{13}^2 & \Delta_{23}^2 & \Delta_{33}^2
\end{array} \right)
+
\left( \begin{array}{ccc}
\tan \beta & -1 & -\frac{\nu_2}{x} \\
-1 & \cot \beta & -\frac{\nu_1}{x} \\
-\frac{\nu_2}{x} & -\frac{\nu_1}{x} & \frac{\nu_1 \nu_2}{x^2}
\end{array} \right)
\Delta_p^2.
\label{blob1}
\ee
The $\Delta_{ij}^2$ and $\Delta_p^2$ are given by
\begin{eqnarray}
\Delta_p^2    & = & \frac{3}{16\pi^2} h_t^2.(\lambda x).A_t.
                    f(m_{t_1}^2,m_{t_2}^2), \nonumber \\
\Delta_{11}^2 & = & \frac{3}{8\pi^2} h_t^4 \nu_2^2.(\lambda x)^2.
                    \left(
                    \frac{A_t+\lambda x \cot \beta}{m_{t_2}^2-m_{t_1}^2}
                    \right)^2
                    g(m_{t_1}^2,m_{t_2}^2), \nonumber \\
\Delta_{22}^2 & = & \frac{3}{8\pi^2} h_t^4 \nu_2^2.\left(
                    \ln \frac{m_{t_1}^2 m_{t_2}^2}{M_{SUSY}^4} +
                    \frac{2A_t(A_t+\lambda x \cot \beta)}
                         {m_{t_2}^2-m_{t_1}^2}
                    \ln \frac{m_{t_2}^2}{m_{t_1}^2} \right) \nonumber \\
              & + & \frac{3}{8\pi^2} h_t^4 \nu_2^2.
                    \left(
                    \frac{A_t(A_t+\lambda x \cot \beta)}
                         {m_{t_2}^2-m_{t_1}^2}
                    \right)^2
                    g(m_{t_1}^2,m_{t_2}^2), \nonumber \\
\Delta_{33}^2 & = & \frac{3}{8\pi^2} h_t^4 \nu_2^2.(\lambda \nu_1)^2.
                    \left( \frac{A_t+\lambda x \cot \beta}
                                {m_{t_2}^2-m_{t_1}^2} \right)^2
                    g(m_{t_1}^2,m_{t_2}^2), \nonumber \\
\Delta_{12}^2 & = & \frac{3}{8\pi^2} h_t^4 \nu_2^2.(\lambda x).
                    \left(
                    \frac{A_t+\lambda x \cot \beta}
                         {m_{t_2}^2-m_{t_1}^2}
                    \right)
                    \left( \ln \frac{m_{t_2}^2}{m_{t_1}^2} +
                    \frac{A_t(A_t+\lambda x \cot \beta)}
                         {m_{t_2}^2-m_{t_1}^2}
                    g(m_{t_1}^2,m_{t_2}^2) \right), \nonumber \\
\Delta_{13}^2 & = & \frac{3}{8\pi^2} h_t^4 \nu_2^2.(\lambda x).(\lambda \nu_1).
                    \left(
                    \frac{A_t+\lambda x \cot \beta}
                         {m_{t_2}^2-m_{t_1}^2}
                    \right)^2
                    g(m_{t_1}^2,m_{t_2}^2) \nonumber \\
              & - & \frac{3}{8\pi^2} h_t^2.(\lambda x).(\lambda \nu_1).
                    f(m_{t_1}^2,m_{t_2}^2), \nonumber \\
\Delta_{23}^2 & = & \frac{\nu_1}{x} \Delta_{12}^2,
\label{blob2}
\end{eqnarray}
where the functions $f$ and $g$ are defined by
\begin{eqnarray}
f(m_{t_1}^2,m_{t_2}^2) & = & \frac{1}{m_{t_2}^2-m_{t_1}^2} \left[
                             m_{t_1}^2 \ln \frac{m_{t_1}^2}{M_{SUSY}^2} -
m_{t_1}^2
                           - m_{t_2}^2 \ln \frac{m_{t_2}^2}{M_{SUSY}^2} +
m_{t_2}^2
                             \right], \nonumber \\
g(m_{t_1}^2,m_{t_2}^2) & = & \frac{1}{m_{t_1}^2-m_{t_2}^2} \left[
                           (m_{t_1}^2+m_{t_2}^2) \ln
\frac{m_{t_2}^2}{m_{t_1}^2}
                           +2(m_{t_1}^2-m_{t_2}^2) \right] .
\end{eqnarray}
If the squarks are degenerate in mass, then Eq.(10)
implies that $A_t+\lambda x \cot \beta =0$. Furthermore, if
$m_{t_1} = m_{t_2} = M_{SUSY}$,
then Eqs.(12-14) imply that the squark contribution
to the CP-even mass--squared matrix vanishes. This is the limit of our
previous analysis in Section 2 \cite{elliott1}.

The CP-odd (pseudoscalar) mass--squared matrix is
\be
M^2_p = \tilde{M}^2 + \delta \tilde{M}^2,
\ee
where $\tilde{M}^2$ is given in Eq.(7) and
\be
\delta \tilde{M}^2 =
\left( \begin{array}{ccc}
\tan \beta & 1 & \frac{\nu_2}{x} \\
1 & \cot \beta & \frac{\nu_1}{x} \\
\frac{\nu_2}{x} & \frac{\nu_1}{x} & \frac{\nu_1 \nu_2}{x^2}
\end{array} \right) \Delta_p^2.
\ee

Finally, the charged mass--squared matrix is
\be
M_c^2 =
\left( \begin{array}{cc}
\tan \beta & 1 \\
1 & \cot \beta
\end{array} \right)
(m_4x-\lambda_7x^2-\lambda_4 \nu_1 \nu_2 + \Delta_c^2),
\label{chargematrix}
\ee
where
\be
\Delta_c^2 = \frac{3}{16\pi^2} \sum_{m_a \in \{ m_{t_1}, m_{t_2}, m_{b_1} \}
}
m_a^2 \left( \ln \frac{m_a^2}{M_{SUSY}^2} - 1 \right)
\frac{\partial^2 m_a^2}{\partial H_1^- \partial H_2^+} \left|_{vevs} \right.
,
\ee
and
\begin{eqnarray}
\frac{\partial^2 m_{t_1}^2}{\partial H_1^- \partial H_2^+}
\left|_{vevs} \right. & = &
-\frac{h_t^4 \nu_2^2.(\lambda x)^2. \cot \beta}
      {(m_{t_1}^2-m_{t_2}^2)(m_{t_1}^2-m_{b_1}^2)}
-\frac{h_t^2.(\lambda x).A_t}{m_{t_1}^2-m_{t_2}^2}, \nonumber \\
\frac{\partial^2 m_{t_2}^2}{\partial H_1^- \partial H_2^+}
\left|_{vevs} \right. & = &
-\frac{h_t^4 \nu_2^2.(\lambda x)^2. \cot \beta}
      {(m_{t_2}^2-m_{b_1}^2)(m_{t_2}^2-m_{t_1}^2)}
+\frac{h_t^2.(\lambda x).A_t}{m_{t_1}^2-m_{t_2}^2}, \nonumber \\
\frac{\partial^2 m_{b_1}^2}{\partial H_1^- \partial H_2^+}
\left|_{vevs} \right. & = &
-\frac{h_t^4 \nu_2^2.(\lambda x)^2. \cot \beta}
      {(m_{b_1}^2-m_{t_1}^2)(m_{b_1}^2-m_{t_2}^2)}.
\end{eqnarray}

The bound on the lightest CP-even Higgs mass is a consequence of the fact
that the minimum eigenvalue of $M_s^2$ is bounded by the minimum eigenvalue
of the upper $2 \times 2$ submatrix of $M_s^2$. Using Eq.~(\ref{chargematrix})
to obtain the physical charged Higgs mass squared, $m_c^2$, in terms of
$m_4$,
\be
m_4x-\lambda_7x^2 = \frac{1}{2} (m_c^2+\lambda_4 \nu^2) \sin 2\beta -
\Delta_c^2,
\ee
we may eliminate $m_4$ from $M_s^2$ in favour of $m_c^2$, and thus write
the upper $2 \times 2$ submatrix in the form
\be
M_s'^2 = M'^2 + \delta M'^2,
\ee
where
\be
M'^2 = \left( \begin{array}{cc}
2\lambda_1\nu_1^2 & 2(\lambda_3+\lambda_4)\nu_1\nu_2 \\
2(\lambda_3+\lambda_4)\nu_1\nu_2 & 2\lambda_2\nu_2^2
\end{array} \right)
+
\left( \begin{array}{cc}
\tan \beta & -1 \\
-1 & \cot \beta
\end{array} \right)
\frac{1}{2}(m_c^2+\lambda_4\nu^2)\sin 2\beta,
\label{treematrix}
\ee
and
\be
\delta M'^2 = \left( \begin{array}{cc}
\Delta_{11}^2 & \Delta_{12}^2 \\
\Delta_{12}^2 & \Delta_{22}^2
\end{array} \right)
+
\left( \begin{array}{cc}
\tan \beta & -1 \\
-1 & \cot \beta
\end{array} \right)
(\Delta_p^2 - \Delta_c^2).
\label{reduceparameters}
\ee
It is clear from the form of the second term on the right--hand--side of
Eq.~(\ref{reduceparameters}) that the factor $\Delta_p^2 - \Delta_c^2$
does not change the bound at all, but merely serves, for a fixed $\tan \beta$,
to shift the bound, when plotted as a function of $m_c$, to the left or right.
We may then drop this term, since its presence may be re-parametrised by
a shift in the free parameter $m_c$. In this way $m_{b_1}$ is
eliminated.

{}From Eq.~(\ref{treematrix}) and Eq.~(\ref{reduceparameters}) it is a simple
matter to determine the shift in the minimum eigenvalue
of $M'^2$ due to squark effects. Defining $A = (M'^2)_{11}$,
$B = (M'^2)_{12}$ and $C = (M'^2)_{22}$, the shift is given by
\be
\frac{1}{2} \left[ \Delta^2_{11} + \Delta^2_{22} -
\frac{(A - C)(\Delta^2_{11} - \Delta^2_{22}) + 4 B \Delta^2_{12}}
     {\sqrt{(A - C)^2 + 4 B^2}} \right].
\label{squarkshift}
\ee
Because this shift is a one loop effect, the couplings $\lambda_i$ in $A$,
$B$ and $C$ may be evaluated at any renormalisation point, since the
difference between a coupling evaluated at two different renormalisation
points is also a one loop effect, thus giving an overall error at the
two loop level, which we neglect in our approximation.

In our previous analysis \cite{elliott1} we used triviality limits on the
couplings $h_t$, $\lambda$ and $k$ \cite{13,14}
to determine the bound on the lightest CP-even
Higgs boson mass $m_h^0$ not including general squark effects, that is, with
$\delta M'^2 \equiv 0$. Let the triviality limit on $\lambda$ for a given
$h_t$ be $\lambda_{max}$. In table \ref{table2} we reproduce the
values of $h_t$ and $\lambda_{max}$ which generated the bound $m_h^0$
of our previous analysis --- all other values, for a
given top quark mass, resulted in smaller values of the bound. It transpires
that $k = 0$ in all cases.
\footnote{Taking $k=0$ gives rise to an axion in the physical spectrum.
However, a small, non--zero value of $k$ is sufficient to give the
would--be axion a mass of several tens of GeV. Such a value of $k$ does not
invalidate our calculations.}
In order to maximise the bound resulting from
Eq.~(\ref{treematrix}) we take the limit $m_c = \infty$, thus eliminating
$m_c$ as a parameter. As a function of $m_c$ the bound approaches its
maximum asymptotically as $m_c \rightarrow \infty$. The approach is rapid,
with $m_c \sim 200$ GeV being a good approximation to $m_c = \infty$.
Thus, the bound resulting from the purely formal procedure of taking the
limit $m_c = \infty$ is not unphysical \cite{elliott1}. Had this not been the
case, then certainly we would have an upper bound, but one perhaps not
capable of realisation in an actual spectrum.

\begin{table}
\caption{\tenrm \baselineskip=12pt
Lightest CP-even Higgs mass bound in the NMSSM including general squark
effects. In row 1 is the top mass, $m_t$, in GeV; in rows 2 and 3 those
values of $h_t(M_{SUSY})$ and ${\lambda}_{max}(M_{SUSY})$,
which produce the bound in our previous analysis, this being in row 4, in
GeV; in row 5 is the contribution to the bound resulting from a general squark
mass spectrum rather than degenerate squarks with mass $M_{SUSY}$, in GeV;
in row 6 is the lightest CP-even Higgs mass bound including general squark
effects, in GeV; in rows 7 and 8 are
those value of $m_{t_1}$ and $A_t$, in GeV, which generate the bound in
row 6; $m_{t_2} = 1$ TeV.}
\begin{center}
\begin{tabular}{|c|r|r|r|r|r|r|r|r|r|r|r|}
\hline
$m_t$ & 90 & 100 & 110 & 120 & 130 & 140 & 150 & 160 & 170 & 180 & 190 \\
\hline
$h_t$ & 0.60 & 0.67 & 0.73 & 0.79 & 0.84 & 0.88 & 0.92 & 0.95 & 0.98 & 1.00 &
1.03 \\
\hline
${\lambda}_{max}$ & 0.87 & 0.85 & 0.83 & 0.81 & 0.79 & 0.77 & 0.74 & 0.71 &
0.67 & 0.63 & 0.50 \\
\hline
$m_h^0$      & 145 & 143 & 140 & 137 & 135 & 132 & 129 & 126 & 124 & 123 & 126
\\
\hline
$\delta m_h$ &   1 &   2 &   3 &   4 &   6 &   8 &  10 &  13 &  16 &  20 &  23
\\
\hline
$m_h$        & 146 & 145 & 143 & 141 & 140 & 139 & 139 & 139 & 140 & 143 & 149
\\
\hline
$m_{t_1}$ & 800 & 770 & 750 & 720 & 700 & 670 & 650 & 630 & 610 & 590 & 570 \\
\hline
$A_t$ & 940 & 1000 & 1040 & 1130 & 1160 & 1270 & 1330 & 1410 & 1500 & 1630 &
1820 \\
\hline
\end{tabular}
\end{center}
\label{table2}
\end{table}

We use the values of $h_t$ and $\lambda_{max}$ in table \ref{table2} to
calculate the shift in the bound on the lightest CP-even Higgs
boson mass--squared given by Eq.~(\ref{squarkshift}). The shift in the
bound on the mass is denoted by $\delta m_h$. This is a hideously
complicated function of many parameters, including the squark masses
$m_{t_1}$ and $m_{t_2}$, the soft SUSY breaking parameter $A_t$,
the vev of the gauge singlet field $x$, and $m_c$. As a function
of $m_c$ the shift appears, numerically, to be maximised for $m_c$ small,
typically between 100 GeV and 200 GeV, with the difference between the shifts
at these two points being less than 1 GeV. Thus, we set $m_c = 200$ GeV in the
general squark contributions to the bound. Doing this enables us to retain
the bound from our previous analysis derived
from Eq.~(\ref{treematrix}) in the limit $m_c = \infty$, and simply maximise
the squark contributions separately and add them to our previous bound
to yield the new bound $m_h = m_h^0 + \delta m_h$.
The $x$ dependence is not too strong, with the ratio
$r = x / \nu$ typically taking values close to 10 in order to maximise the
bound. Where this is not the case, the difference between the bound at its
maximum, as a function of $x$, and that at $r=10$ is less than 1 GeV.
Thus, we set $r=10$, this giving a reliable indication of the maximum shift.
We calculate the shift for various ranges of squark masses and a range of
values of $A_t$, and record the maximum value obtained. The squark masses
are allowed to vary between 20 GeV and 1000 GeV in steps of 10 GeV. (1000 GeV
is the upper limit since we take $M_{SUSY} = 1000$ GeV.) Given this
restriction, the value of $A_t$ which maximises the general squark
contribution to the bound never exceeds 2 TeV. We impose the constraint
$ 2h_t \nu_2 (A_t + \lambda x \cot \beta) \leq |m_{t_1}^2-m_{t_2}^2|$ which
follows from the form of the squark mass matrix in Eq.~(\ref{squarkmatrix}).

In table \ref{table2} we show the bound on the lightest CP-even
Higgs mass $m_h$ including general squark effects (and $m_h^0$ for
comparison). Table \ref{table2} also shows the values of $m_{t_1}$ and $A_t$
used to generate the bound.
As $m_t$ increases, $m_{t_1}$ monotonically decreases, while $A_t$
monotonically increases. The other squark mass, $m_{t_2}$, takes the value
1 TeV over the whole range of top quark masses --- that one squark mass takes
on its maximum permitted value to maximise the general squark contribution to
the bound can be seen analytically. It will be noticed that these squark
masses and $A_t$ are somewhat larger than typically expected from GUT
scenarios. However, we feel it wise not to restrict ourselves
too excessively to particular prejudices regarding physics as yet unknown
(though, of course, the notion of triviality does require the assumption
of a SUSY desert up to the unification scale).

Our calculations suggest that the universal
upper bound on the lightest CP-even Higgs mass is 149 GeV.
Table \ref{table2} reveals that for
large values of $m_t$ squark effects may contribute
up to $\delta m_h=23$ GeV, but for small $m_t$
the squark effects are small, as expected. We emphasise that we have not
re-maximised the complete one loop corrected bound, but rather made the
assumption that the values of $h_t$ and $\lambda_{max}$ which generated the
bound of our previous analysis are not significantly modified by the
inclusion of a general spectrum of squark masses.

\vglue 0.5cm

{\bf \noindent 4. Conclusions \hfil}
\vglue 0.4cm

We have combined an RG analysis of the Higgs
sector of the NMSSM \cite{elliott1}
with a calculation of general squark effects \cite{elliott2}, and
determined a bound on the mass of the lightest CP-even
state of $m_h\leq 146,139,149 \ GeV$ for $m_t=90,140,190 \ GeV$.
Our calculations indicate that the effects of a general squark
spectrum can be very significant for large top quark masses. One reason why,
for large top quark masses, squark effects
are important is that there exist finite, one--loop diagrams with vertices
containing the uncontrolled soft SUSY breaking parameter $A_t$. These
diagrams can give large contributions to the Higgs boson masses. We close
by indicating that in the NMSSM there are similar diagrams involving loops
of Higgs bosons, again with vertices containing soft SUSY breaking parameters,
this time $A_{\lambda}$ and $A_k$. The existence of these diagrams is
directly due to the gauge singlet field $N$; they do not occur in the MSSM.
We see no reason in principle why these new diagrams should not give rise
to similarly large effects, except for obvious factors of 3 due to colour.
We are currently in the process of estimating these effects \cite{elliott3}.

\vglue 1.0cm

\begin{center} {\bf Acknowledgements} \end{center}
This work was done in collaboration with Terry Elliott and Peter White,
at the University of Southampton.
I would like to thank the organisers of the workshop for their
hospitality and the SERC for financial support.

\newpage
\bibliographystyle{unsrt}

\end{document}